# EA-ERT: a new ensemble approach to convert time-lapse ERT data to soil water content

**B. Loiseau[1,2], S. D. Carrière[1,2], N. K. Martin-StPaul[3], R. Clément[4], C. Champollion[5], V. Mercier[6], J. Thiesson[1], S. Pasquet[1,7], C. Doussan[6], T. Hermans[8], D. Jougnot[1]**

[1]UMR METIS, Sorbonne Université, UPMC, CNRS, EPHE, 75005 Paris, France.

[2] HSM, Univ. Montpellier, CNRS, IMT, IRD, Montpellier,

[3]URFM, INRAE, Domaine Saint Paul, Site Agroparc, 84000 Avignon, France.

[4]REVERSAAL Research Unit, INRAE, Villeurbanne, 69626, France.

[5]Géosciences Montpellier, Université de Montpellier, CNRS, Montpellier, France.

[6]INRAE - Avignon Université, UMR EMMAH, 84914, Avignon, France.

[7]Observatoire des Sciences de l'Univers, ECCE TERRA, UAR 3455, CNRS, Sorbonne Université, Paris, France

[8]Department of Geology, Ghent University, 9000 Gent, Belgium.

Corresponding author: Bertille Loiseau ([bertille.loiseau@ird.fr](mailto:bertille.loiseau@ird.fr))

**Key Points:**

- An ensemble approach to process and convert ERT data to water content.

- The method circumvents inversion parameter choice issues.

- The method estimates a form of uncertainty in the final model to assess reliability.

## Abstract

Electrical Resistivity Tomography (ERT) is increasingly used to study subsurface hydrological processes. It shows promising potential for estimating soil water content, a key but challenging property to quantify. However, converting the resistivity signal into water content is complex. This encourages developing approaches to increase the robustness of estimates while facilitating the evaluation of uncertainties. In this paper, we propose an innovative method, called the Ensemble Approach ERT (EA-ERT), to build an ensemble model of electrical resistivity calibrated from field data and then to convert it into a spatial distribution of water content. This approach combines time-lapse ERT data with point-based in-situ soil water content measurements. It enables i) circumventing inversion parameter choice by evaluating the performance of a large number of models, ii) estimating uncertainty in the final model by calculating the coefficient of variation among the models composing the ensemble, and iii) converting electrical resistivity models to water content. The method was tested at two dissimilar field sites in southern France. For each site, an ensemble model, built from multiple inversions, was selected and converted into soil water content. The calculated values showed a good fit (r ≥ 0.8), with small differences (RMSE ≤ 3.24 % vol.) compared to in-situ measurements. Areas of high uncertainty were identified, providing complementary information to the more classical indicators from the inversion code. EA-ERT provides a robust and automatable method to convert ERT data to related parameters, contributing to improved monitoring and understanding processes in the subsurface.

**Plain Language Summary**

Soil water content is an essential but difficult property to estimate. Electrical Resistivity Tomography (ERT) is an increasingly used technique to study it. ERT is a geophysical technique that measures how much the ground resists the flow of electrical current. This resistance is related to the amount of water in the soil. Field measurements are processed using specialized software to produce underground resistivity image. However, converting the resistivity signal into water content estimates is complex, because the relationship is not linear and depends on local soil conditions. This has led researchers to develop methods that make estimates more reliable and better quantify uncertainty. In this study, we propose an innovative method, the Ensemble Approach ERT (EA-ERT), which processes ERT data and converts it to water content. The method combines several resistivity images that match well with field water content measurements to produce a more accurate subsurface image. This ensemble approach helps test different processing choices, estimate uncertainty by calculating variability among combined images, and convert the result into a water content map. EA-ERT is a robust, automatable method that allows ERT data to be converted into useful parameters such as water content, improving monitoring and understanding of subsurface processes.

# 1. Introduction

Soil water content is a key property in many domains such as hydrology, agronomy, geology, ecology, biology and environmental studies. In particular, it is essential for understanding and modelling hydrological and ecohydrological processes within the critical zone (e.g., Robinson et al., 2008a; Vereecken et al., 2008; Jonard et al., 2018). It plays an important role in the water cycle by regulating the rainfall–runoff response and determining the soil water availability for evapotranspiration (e.g., Robinson et al., 2008a; Jonard et al., 2018). However, estimating soil water content is challenging due to the variability of soils properties such as structure, texture, composition (e.g., Samouelian, 2005; Vereecken et al., 2022), and is further complicated by its spatial and temporal variability.

The reference method for estimating the subsurface water content is gravimetric measurement on soil samples, which is labor-intensive and destructive. A large number of indirect methods have been developed, and several reviews have described them (e.g., Gardner et al., 2000 provided an initial overview with a focus on the physical principles and practical challenges; Robinson et al., 2008a and Vereecken et al., 2008 emphasized data integration for field- and catchment-scale hydrological studies; S.U. et al., 2014 offered a critical evaluation of various techniques; Jonard et al., 2018 focused on ground-based methods, particularly field-scale hydrogeophysics). The most common methods are point-scale methods that involve inserting sensors into the soil at different depths, such as time-domain reflectometry (TDR), frequency-domain reflectometry (FDR), and neutron probes. These methods offer high temporal resolution but provide limited spatial representativness, often limited in depth. Field-scale methods are being developed to map spatial and temporal water content variations with good resolution. Among them, hydrogeophysical techniques such as ground-penetrating radar (GPR), electromagnetic induction (EMI), and electrical resistivity tomography (ERT) are increasingly used to estimate the water content (e.g., Huisman et al., 2003; Robinson et al., 2008b; Binley et al., 2015; Jonard et al., 2018; Hermans et al., 2023). These methods are non-destructive, cover large areas, and offering integrative measurements of subsurface properties.

ERT is a widely used technique for investigating subsurface water distribution and its temporal dynamics by mapping the subsurface electrical resistivity, which is dependent, among other factors, on the water content (e.g., Daily et al., 1992; Parsekian et al., 2015; Slater and Binley, 2020). ERT involves injecting electrical current into the ground and measuring the resulting potential differences, typically at the surface, although borehole applications are possible. These measurements are used to calculate apparent resistivity values, which reflect the integrated response of the subsurface resistivity distribution beneath each measurement quadrupole (e.g., Loke, 1999). The spatial distribution of subsurface resistivity is then reconstructed through an inversion process, which makes it possible to image 2D sections or 3D volumes of the subsurface (e.g., Binley and Kemna, 2005). ERT popularity stems from its ease of deployment, good spatial and depth coverage, and its ability to repeat measurements for time-lapse monitoring. Because electrical resistivity is strongly influenced by soil water content, ERT is particularly suited for hydrological applications. Time-lapse ERT and its conversion to water content have been used to characterize water infiltration pathways and root water uptake zones (e.g., Daily et al., 1992; Michot et al., 2003; Cassiani et al., 2015; Sakar et al., 2026), assess water competition between different plant species (e.g., Garré et al., 2013), and analyse soil-plant interactions under various environmental conditions (e.g., Beff et al., 2013; Thayer et al., 2018; Dafflon et al., 2023).

However, it is well known that estimating water content from apparent resistivity data can lead to significant uncertainties, largely due to (1) the inversion process and (2) the adjustment of the petrophysical relationships between the two variables (e.g., Day-Lewis et al., 2005; Linde et al., 2017; Jougnot et al., 2018). Firstly, inversion is an ill-posed problem, meaning that there are many resistivity models that can explain the measured data (e.g., Binley and Slater, 2020). Deterministic gradient-based inversion methods are the most widely used because of their speed and stability (e.g., Loke et al., 2003; Günther, 2004; Rücker et al., 2017; Blanchy et al., 2020; Binley and Slater, 2021). These methods rely on regularization to stabilize the solution. However this often results in smoothing that may reduce contrast or mask certain subsurface heterogeneities (e.g., Zhou et al., 2014). Moreover, despite efforts to democratize access to geophysical inversion tools through simplified programs (e.g., Loke, 2001; Blanchy et al., 2020) the inversion process still requires expertise to correctly parameterize the models and avoid artifact misinterpretation (e.g., Clément et al., 2009, 2010; Carrière et al., 2013; Simyrdanis et al., 2015).

In contrast, probabilistic approaches—such as Bayesian methods or Markov Chain Monte Carlo algorithms—allow for quantification of uncertainties, but they are computationally intensive and require reliable prior information (e.g., Hermans et al., 2016; Galetti and Curtis, 2018). Some alternative approaches aim to reduce this computer workload, such as the Ensemble-Based ERT method proposed by Aleardi et al. (2021), which combines numerical efficiency with uncertainty exploration. Recently, Arboleda-Zapata et al. (2025) applied this approach to time-lapse ERT to monitor water flow during managed aquifer recharge, demonstrating its ability to capture wetting front dynamics and related hydrogeological processes.

Time-lapse ERT inversion introduces additional challenges related to the comparison of models at different time steps. The observed differences may reflect actual subsurface changes such as variations in water content but can also arise from artifacts introduced by noise, regularization, or the inversion algorithm itself (e.g., Binley and Kemna, 2005; Miller et al., 2008). The most common approach consists in inverting each dataset independently and then analyzing the resulting resistivity differences. However, this method can amplify non-physical artifacts and lead to temporal inconsistencies between models (e.g., Loke, 1999; Clément et al., 2009). To address these limitations, alternative inversion strategies have been proposed, including the use of a common reference model or cascade (sequential) inversion, where the result from one time step serves as the reference model for the next (e.g., Loke, 1999; Miller et al., 2008). These approaches improve the ability to track subtle temporal variations and enhance inversion stability, but they remain sensitive to parameter choices and systematic errors (e.g., Karaoulis et al., 2014).

To reduce uncertainties in the inverse models, information fusion methods that combine multiple data sources to guide interpretation or improve decision-making have been explored (e.g., Dezert, 2019; Rabouli et al., 2021; Sakar et al., 2024). However, these methods require prior knowledge of uncertainties and sufficiently rich datasets, which are rarely available in practice. Another solution is to use joint inversion (e.g., Linde and Doetsch, 2016; Mollaret et al., 2020; Chen and Niu, 2022), which integrates complementary data to better constrain subsurface structure and reduce uncertainty. A more advanced alternative is coupled inversion, which explicitly solves flow equations to link hydrological and geophysical parameters (e.g., Hinnell et al., 2010). However, this approach is too demanding if the main interest lies in water content distribution. Both approaches require careful data parameterization and model design.

Some metrics have been used to assess the reliability of inverted models, but results have not always been satisfactory. Reliability is generally assessed by calculating error criteria (e.g., RMS, Chi²) that estimate the mathematical difference between measured and simulated apparent resistivities. These criteria can be used to assess the quality of one inversion compared with another, but due to the non-uniqueness of the inversion, several inversions can have the same error criterion (e.g., Caterina et al., 2014), and the inversion with the lowest error criterion does not always yield the model most similar to field reality (e.g., Descloitres et al., 2008). Other parameters have been developed to express model reliability spatially along the profile, such as resolution matrix (e.g., Friedel, 2003; Oldenborger and Routh, 2009), sensitivity (e.g., Robert et al., 2012), cumulative sensitivity (e.g., Günther, 2004; Nguyen et al., 2009), depth of investigation index (DOI), (Oldenburg and Li, 1999; Caterina et al., 2013; Carrière et al., 2017) and uncertainty (e.g., Loke, 2001). These criteria are powerful tools designed to provide smooth, depth-dependent information about model reliability, particularly useful for assessing the depth of investigation (e.g., Caterina et al., 2013; Paepen et al., 2022). Therefore, since they were not designed to detect localized anomalies, they are generally not sensitive to point artifacts and do not fully quantify spatial uncertainty.

Adjusting the relationship between electrical resistivity and water content is a complex task as it is non-linear and depends on soil properties (e.g., Friedman, 2005; Samouëlian et al., 2005; Glover, 2015). A misestimation of the relationship can lead to significant errors in water content estimation (e.g., Tso et al., 2019). This is especially true in heterogeneous environments, where a single relationship cannot be applied across the entire soil profile (e.g., Brillante et al., 2015). A review by Brillante et al. (2015) examined the methods used to develop models enabling the use of ERT to spatialize water content. These methods fall into two categories: (1) experimental calibrations using regression analysis with paired measurements of water content and electrical resistivity (e.g., Michot et al., 2003; Calamita et al., 2012; Brillante et al., 2014), and (2) the use of petrophysical models based on the soil physical properties (e.g., Brunet et al., 2010; Garré et al., 2011; Beff et al., 2013; Thayer et al., 2018; Michot et al., 2020). Several models have been proposed to describe the relationship between electrical resistivity and water content (e.g., Friedman, 2005; Samouëlian, 2005), notably the well-known Waxman and Smits (1968) model which can be seen as an extension of Archie (1942) for media where the surface conductivity cannot be neglected (i.e., typically soils, see Doussan and Ruy, 2009). Regardless of the method, these relationships can be established either in the laboratory on soil samples (e.g., Zhou et al., 2001; Brunet et al., 2010; Chen and Niu, 2022) or directly in the field using available acquired data (e.g., Michot et al., 2003; Beff et al., 2013; Garré et al., 2011). Laboratory calibration is simpler and faster, but its application in the field is debated due to the disturbance of soil structure during sampling and scaling issues (e.g., Michot et al., 2003; Brillante et al., 2015; Dimech et al., 2023). Field calibration is more robust, but it requires many point-scale measurements to obtain sufficient variation in water content to properly adjust the relations (e.g., Brillante et al., 2015). Moreover, as it relies on the inverted resistivity model, this calibration includes uncertainty from the inversion step itself (e.g., Hermans and Irving, 2017) and can be spatially dependent due to the variable sensitivity of the inversion (e.g., Day-Lewis et al., 2005).

Therefore, the estimation of water content from ERT data raises three major challenges: (1) the technical aspect of the inversion process and in particular the choice of inversion parameters based on prior knowledge about the medium and on the literature (e.g., Loke et al., 2003, Audebert et al., 2014), (2) the uncertainty and sensitivity of the inverted resistivity model to ensure correct interpretation of the data (e.g., Oldenburg and Li, 1999), and (3) the conversion of electrical resistivity into parameters directly useful for practitioners, in our case soil water content (e.g., Friedman et al., 2005; Samouëlian et al., 2005; Laloy et al., 2011).

In this paper, we propose a new method called Ensemble Approach ERT (EA-ERT) for processing and converting ERT data to water content. The method aims to estimate the subsurface water content distribution and its associated uncertainties by combining (i) time-lapse ERT data (ii) and point-scale in situ soil water content measurements. EA-ERT is an ensemble approach inspired by techniques that are widely used in spatial hydrology to estimate precipitation (e.g., Beck et al., 2017; Ollivier et al., 2023) or evapotranspiration (e.g., Allies et al., 2020). The methodology is detailed in Section 2. It is applied to two dissimilar field areas — one a wheat field on a fairly homogeneous clay-loam soil and the other a meadow located adjacent to a stand of black pine trees on a heterogeneous dolomitic soil atop a karstic system — and the results are presented in Section 3. In addition to the field applications, the methodology was validated using synthetic examples presented in the Supplementary Information. The contributions and limitations of the EA-ERT method are discussed in Section 4, before concluding in Section 5.

## 2. A new methodology for processing ERT data

The proposed method for building a subsurface resistivity model is based on an analysis and combination of several inversions. The methodology is divided into five main steps (Figure 1):

- **STEP 1: Generate inverted models** – The user chooses the inversion parameters and their values to be tested, and uses each parameter set to invert the time-lapse ERT field data, resulting in a total of $N$ time-lapse inverted models.

- **STEP 2: Calibrate petrophysical relationship** – For each of the $N$ inverted models, estimated electrical resistivity values collocated with water content observations are extracted. Specifically, observation pairs $[\rho(x_i, t_j), \theta(x_i, t_j)]$ are extracted from locations $x_i$ and survey times $t_j$. An independent petrophysical relationships is fitted for each of the extracted observation sets and an *RMSE* value between predicted and observed water content is calculated for each of the $N$ inverted models.

- **STEP 3: Inverted model averaging** – The set of $N$ inverted models is ranked ($r$) in order of increasing *RMSE*, i.e., the inversion producing the smallest water content error is ranked $r = 1$ and the one with the largest error is $r = N$. Models are then combined using a cumulative weighted average, i.e., $\bar{\rho}_r = \sum_{i=1}^{r} w_i \rho_i$, weighted by $w_i$ such that $r = 1$ returns the best fitting model, $r = 2$ returns the average of the two best fitting models, $r = 3$ returns the average of the three best fitting models, and so on until $r = N$ is the average of all $N$ inverted models generated. The goal of this procedure is to gradually average out artifacts in the ERT inversions that are inconsistent with the water content data.

- **STEP 4**: **Recalibrate petrophysical relationship** – The process described in Step 2 to extract data pairs and fit petrophysical relationships is repeated, but this time using the average inverted models from Step 3 rather than the individual inverted models from Step 1. As in step 2, the error in predicted water content is estimated for each model from $r = 1$ to $N$. It is important to note that it is unlikely that $r = 1$ will have the lowest error.

- **STEP 5: Model selection and assessment** – The averaged model $\bar{\rho}_K$ with the smallest value of *RMSE* is selected as the "best" model and the coefficient of variation (*CV*) at each cell is calculated from the top $= 1$ to $K$ models that were averaged to produce $\bar{\rho}_K$. The goal is to evaluate the reliability of the selected model.

In the following subsections, specific details are provided for each of these steps after describing the field configuration and data requirements necessary for applying this method. The method was first validated through a synthetic study presented in the Supplementary Information (Text S1 and Figures S1 to S4), and two field applications are presented in the section 3, "Field applications".

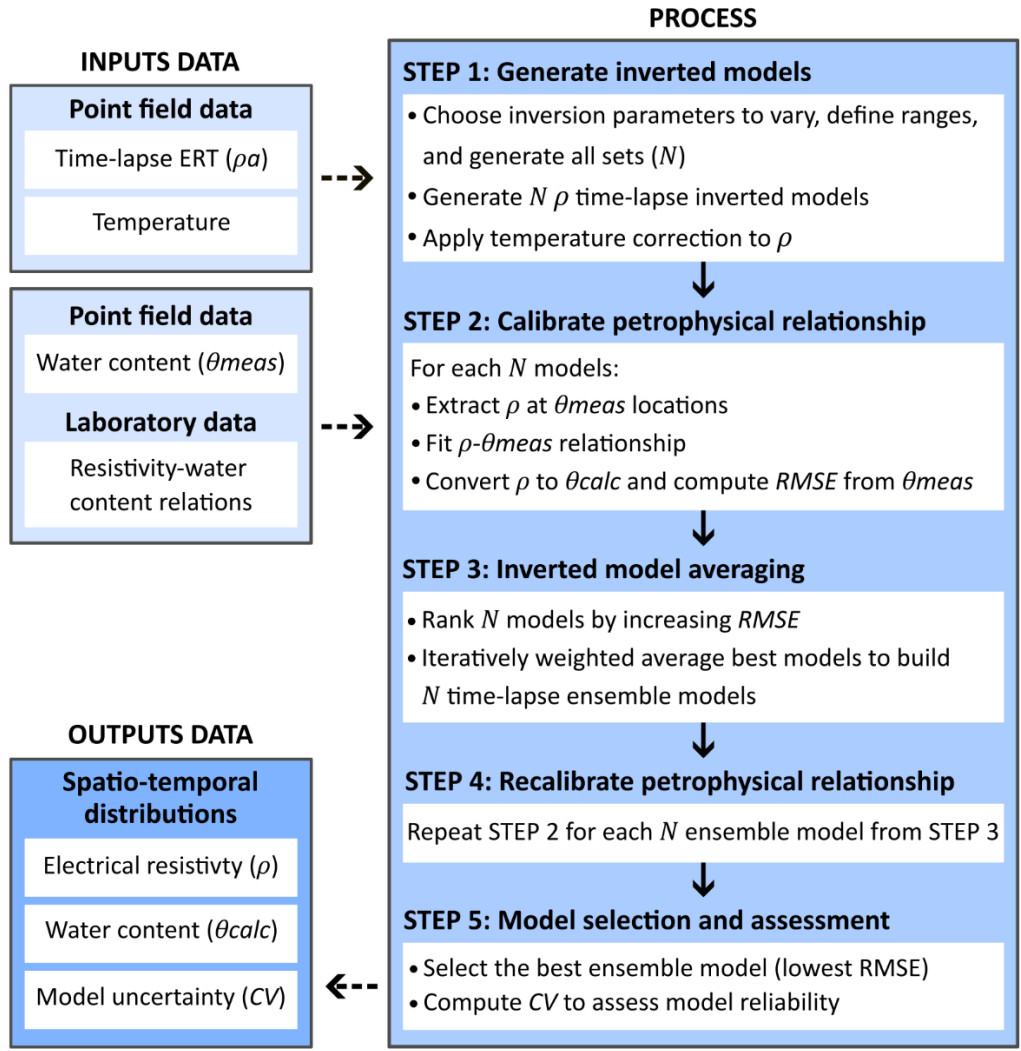

*Figure 1: Flow diagram illustrating the data requirements and the five steps of the EA-ERT method. The required data include: temperature and volumetric water content $\theta_{meas}$ measured at specific points using in-situ sensors, apparent electrical resistivity $\rho_a$ from time-lapse ERT field measurements, and laboratory data (paired measurement of electrical resistivity $\rho_{lab}$ and volumetric water content $\theta_{lab}$) obtained from soil samples (not mandatory). $\rho$ refers to the inverted electrical resistivity and $\theta_{calc}$ is the volumetric water content estimated from $\rho$ using petrophysical relationships. RMSE is the root mean square error. Data requirements and all steps are detailed in sections 2.1 to 2.5.*

## 2.1. Method requirements

This method requires, at a minimum, repeated time-lapse ERT measurements along a single profile, as well as independent point measurements of water content distributed along that profile (Figure 1). Note that in the following, "water content" stands for the volumetric water content $\theta$, related to the water saturation $S_w$ by the porosity $\phi$ : $\theta = S_w * \phi$. The resolution of ERT measurements and those of soil water content probes differ significantly. ERT offers spatial resolution ranging from tens of centimeters to several meters, depending on the spacing between electrodes, whereas water content probes measure over a few centimeters around the probe. Probes are essential for precise point measurements. In contrast, ERT enables the production of continuous spatial mapping of resistivity variations at a scale that probes alone cannot reach.

Electrode spacing along the ERT profile should be chosen to be sensitive to areas where water content probes are installed. For example, a fairly narrow electrode spacing (< 1 m) is necessary to be sensitive to shallow soil horizons. Water content probes (e.g., TDR, FDR, neutron probes) should be installed at various depths in the soil, at least one per soil horizon, with replicates if possible, and at various positions along the ERT profile to account for environmental heterogeneity and for variations in imaging resolution (e.g., Day-Lewis et al., 2005). ERT measurement collected before the actual survey can help optimize probe placement by providing spatial information on soil heterogeneity. Ideally, probes should be installed as deep as possible to properly calibrate the relationship between electrical resistivity and water content, at least down to the levels where variations of interest are expected to occur. It is useful to have water table information to better calibrate the data at depth. Moreover, for the calibration between resistivity and water content data to be meaningful, probe installation must take into account the depth of investigation as well as the resolution of the ERT device used. For instance, if the electrode spacing is two meters, it is advisable to install some water content probes deeper than one meter, since the ERT will mainly be sensitive to deeper volumes.

A sufficiently wide range of water content values is essential to use the method and to properly establish the relationships between resistivity and water content. The more data available and varied, the more robust the relationships established. Therefore, this article presents a time-lapse measurement method that captures data under varying water content conditions to account for this variability. It is also advisable to know the environment of the study area, i.e. soil type and its horizons, in order to check the calibration values obtained by comparing them with literature data or through measurements carried out in the laboratory. In the case of time-lapse measurements, temperature probes are also recommended to correct electrical resistivity measurements for temperature effects over time and with depth (e.g., Hayley et al. 2010). Petrophysical relationships established in the laboratory can be used as a starting model for estimating field-scale relationships (see section 2.3) or for deeper horizons such as bedrock where direct measurements are lacking.

### 2.2. Generate inverted models

The first step of the method (Figure 1) is to generate a large number $N$ of time-lapse inverted models. The method can be used with various inversion software packages (e.g., BERT (Günther and Rücker, 2015), ResIPy (Blanchy et al., 2020), pyGIMLi (Rücker et al., 2017), RES2DINV (Loke et al., 2001)). More generally, any inversion codes that can be automated and that allows users to manage both standard inversion parameters and those specific to time-lapse inversions. First, an appropriate set of inversion parameters that may influence the inversion result must be defined. This involves selecting which parameters will be varied and specifying the range of values to be explored for each of them. This choice should be adapted to the characteristics of the environment under investigation. All possible combinations of these parameter values are then generated to produce $N$ sets of inversion parameters.

For example, using BERT inversion code (Boundless Electrical Resistivity Tomography; Günther and Rücker, 2015), Audebert et al. (2014) identified three inversion parameters that have a major effect on the final inverted model: model norm, anisotropy factor (*zweight*), and regularization parameter ($\lambda$). These parameters are described as follows:

- The model norm corresponds to the model constraint: sharp (*L1*) or smooth (*L2*) boundaries. "L1" is more suitable for abrupt resistivity variations in the subsurface, whereas "L2" is more suitable for progressive variations.
- The anisotropy factor "*zweight*" (i.e. the flatness ratio) defines the weight to be applied to horizontal (< 1) or vertical (> 1) structures.

- The regularization parameter "$\lambda$" (i.e. the damping factor) determines the balance between data fit and model roughness: small values lead to rough models favoring a good data fit, while large values correspond to smoother models favoring the model constraint.

Since the method relies on time-lapse data, the choice of the reference model for the inversion is a fourth important parameter. A standard approach is to invert each dataset independently using a different homogeneous starting model for each time step. To limit inconsistencies and the appearance of artifacts over time, alternative strategies are used for time-lapse inversion. One such strategy is to invert all time steps using a common reference model, often the inverted model from the first time step. In this case, it is possible to use the same or a different "$\lambda$" for time step inversions as was used for the reference model. Another approach is cascade (or sequential) inversion, where the result from one time step serves as the reference model for the next. Each of these strategies presents its own advantages and disadvantages (e.g., Günther et al., 2006; Singha et al., 2015)

Once the $N$ sets of inversion parameters are defined, the time-lapse ERT data are inverted using each set to produce $N$ time-lapse inverted models. The electrical resistivity values ($\rho$) of the models should be corrected for temperature effects, since the study involves time-lapse data. For this, we use the model of Campbell et al. (1948)

$$\rho_{corr} = \frac{\rho}{1 + \alpha(T - Tref)} \tag{1}$$

where $\rho_{corr}$ is the temperature-corrected electrical resistivity, $T$ the soil temperature from point sensors located at different depths and lateral positions, $Tref$ is a chosen reference temperature, and $\alpha$ is the temperature coefficient of resistivity. Note that $\alpha$ can vary slightly depending on the fluid composition and can be determined experimentally (e.g., Hayley et al., 2007; Hermans et al., 2014; Travelletti et al., 2012). In field applications (section 3), $Tref$ = 20°C and $\alpha$ = 0.023°C$^{-1}$ (Travelletti et al., 2012) were used.

### 2.3. (Re)Calibrate petrophysical relationship

The second step of the method (Figure 1) consists of evaluating the $N$ time-lapse inverted model to predict soil water content. First, the inverted electrical resistivity values are extracted at the water content probe locations for each time step and for all $N$ models. It is important to note, however, that ERT and water content measurements differ significantly in terms of spatial resolution and integration volume. ERT integrates over significantly larger volumes (≈ m3) compared to water content sensors (≈ cm3). This difference in the measurement footprint highlights the importance of carefully selecting the integration volume when extracting ERT resistivity values for comparison with point-scale water content measurements (e.g., Benoit et al., 2019; Dimech et al., 2023; Terry et al., 2023). To smooth out lateral anomalies or extreme values, the $\rho$ values are averaged within a horizontal window centered on a sensor location prior to extraction. The width of the horizontal averaging window is chosen as a compromise between retaining variability and damping noise in the image. Tests should be done to find the optimum integration width (which varies from survey to survey). The method can be tested for different integration widths to determine the one that best matches water content estimated from inverted electrical resistivity with measured values. Alternatively, a faster approach is to use a power or exponential law to find the integration width giving the best estimate of water content from electrical resistivity data (e.g., Calamita et al., 2012). For each time-lapse inverted model, the ERT-derived electrical resistivity is paired with the volumetric water content measured by the soil probes (the same measurements are used for all $N$ inversions). An error can be estimated from the standard deviation of inverted electrical resistivity over the integration width.

For the following analysis, the dataset (water content versus electrical resistivity) is split into two independent subsets. One subset is used for training, the other for validation. The training dataset is used to evaluate and combine inverted models, while the validation dataset is used at the end to assess the performance of the final selected ensemble model. Each subset should ideally cover a sufficiently wide range of volumetric water content values, particularly the training dataset, to ensure accurate calibration of the petrophysical relationships. Independence between the two subsets can be ensured, for example, by using data from every other time step, or by assigning measurements from one pit to the training set and measurements from other pits to the validation set.

Next, petrophysical relationships linking electrical conductivity $\sigma = 1/\rho$ (S.m$^{-1}$) to volumetric water content $\theta$ (m$^3$/m$^3$) are fitted using the training dataset. A separate relationship is established for each soil horizon and for each inverted model using the Waxman and Smits (1968) model:

$$\sigma = \phi^m S_w^n \left( \sigma_w + \frac{\sigma_s}{S_w} \right) \tag{2}$$

Where $\phi$ (-) is soil porosity, $m$ (-) cementation exponent, $S_w = \theta/\phi$ (-) water saturation, $n$ (-) saturation exponent, $\sigma_w$ (S.m$^{-1}$) water conductivity, and $\sigma_s$ (S.m$^{-1}$) surface conductivity. $\sigma$ corresponds to the electrical conductivity obtained from ERT data included in the training dataset and $\theta$ corresponds to the water content from soil sensors. $\phi$ is either fixed if known, or optimized by providing limits consistent with the soil horizon under study. The parameters $m$, $n$, $\sigma_w$ and $\sigma_s$ are optimized by minimizing the function $f$ below, where $\sigma_{ERT}$ corresponds to the inverted electrical conductivities extracted from ERT:

$$f = \sum \left| \frac{\log(\sigma_{ERT}) - \log(\sigma)}{\log(\sigma_{ERT})} \right| \tag{3}$$

The logarithms (base 10) of $\sigma_w$ and $\sigma_s$ in S/m are used to constrain positive values. The *optim* function of the R programming language (R Core team 2022) is used by applying the "L-BFGS-B" method (Byrd et al., 1995) to constrain each variable within bounds defined based on the literature and site-specific considerations Laboratory measurements can be done on soil or rock samples collected in the field to determine the initial value of parameters.

Petrophysical relationships are then used to convert the electrical resistivity data extracted from each model, included in the training dataset, into water content. An RMSE (Root Mean Square Error, see Equation 4) value between the calculated volumetric water content ($\theta_{calc}$) and the measurements ($\theta_{meas}$) from the probes is calculated for each of the N inverted models. The RMSE is used to quantify the differences, giving greater weight to large discrepancies.

$$RMSE = \frac{1}{N} \sqrt{\sum_{i=1}^{N} (\theta_{calc} - \theta_{meas})^2} \tag{4}$$

Step 4 follows the same procedure, but is applied to the $N$ time-lapse ensemble models.

### 2.4. Inverted model averaging

The goal of the third step is to construct ensemble models that best explains the water content probe data (Figure 1). First, the inverted models are ranked by increasing RMSE, i.e., the model with the smallest water content error is ranked $r = 1$ and the one with the largest error is $r = N$. Ensemble models are then constructed by cumulatively calculating weighted averages of the individual inverted models, starting from the best-fitting model:

$$\bar{\rho}_r = \sum_{i=1}^{r} w_i \rho_i \qquad (5)$$

Where $\bar{\rho}_r$ is the electrical resistivity of the ensemble model including the r best-fitting models, $\rho_i$ is the electrical resistivity of the $i$-th model, $w_i$ is the weight assigned to the $i$-th model. Consequently, $r = 1$ corresponds to the best fitting model, $r = 2$ to the weighted average of the two best-fitting models, $r = 3$ to the weighted average of the three best-fitting models, and so on until $r = N$ which is the weighted average of all $N$ inverted models.

Weights are determined from each model's RMSE. The RMSE values of all $N$ individual models $i$ are normalized using the maximum ($RMSE_{max}$) and minimum ($RMSE_{min}$) values among the $N$ models:

$$RMSE_{norm,i} = \frac{RMSE_i - RMSE_{max}}{RMSE_{min} - RMSE_{max}} \qquad (6)$$

Then, for each ensemble model, a weight $w_i$ is assigned to each constituent model $i$ based on its normalized RMSE, giving more importance to more reliable models (lower RMSE):

$$w_i = \frac{RMSE_{norm,i}}{\sum_{j=1}^{r} RMSE_{norm,j}} \qquad (7)$$

The weights are recalculated for each ensemble model to ensure that their sum equals 1. This weighting method is similar to the generalized likelihood uncertainty estimation (GLUE) approach of Beven and Binley (1992). The general approach is based on that used by Ollivier et al. (2023) to produce a precipitation data set.

### 2.5. Model selection and assessment

After recalibrating the petrophysical relationships for each ensemble model and calculating the RMSE for all $N$ ensemble models in Step 4 using the training dataset, the ensemble model $\bar{\rho}_K$ with the lowest RMSE is selected in Step 5 as the "best" model. Here, $K$ denotes the number of models averaged in the selected ensemble. It corresponds to the best fit among the ensemble model considered.

Selecting an electrical resistivity model based on the combination of several inverted models makes it possible to quantify the dispersion of resistivity values between averaged models. For each cell $c$ within the 2D section, the weighted standard deviation ($sd_c$) of the weighted mean resistivity values $\bar{\rho}_c$ from the $K$ models is calculated as a dispersion indicator. The coefficient of variation ($CV$) is then computed as:

$$CV_c = \frac{sd_c}{\bar{\rho}_c} * 100 \qquad (8)$$

Where $\bar{\rho}_c = \sum_{i=1}^{K} w_{c,i}\rho_{c,i}$ and $sd_c = \sqrt{\frac{K}{K-1}\left(\sum_{i=1}^{K} w_{c,i}(\rho_{c,i} - \bar{\rho}_{c,i})^2\right)}$ .

The *CV* allows comparison of the relative variability of resistivity values in space (within the 2D section) and over time in time-lapse studies (e.g., MacLeod et al., 2016; Vinciguerra et al., 2024). Expressed as a percentage, high *CV* values (above 100%) indicate that the variability among weighted resistivity values exceeds their weighted mean, reflecting high uncertainty; in such case, the corresponding resistivity and water content should be interpreted with caution.

### 2.6. Validating the ensemble model

In order to validate the selected ensemble model, the validation dataset is used. Electrical resistivity values from this dataset are converted into volumetric water content using petrophysical relationships fitted on the training dataset, applied to the selected ensemble model. The RMSE between the calculated and measured water contents is calculated for validation.

A synthetic data study was conducted to rigorously assess and validate the proposed EA-ERT method prior to its application to field data. Details of this analysis are provided in the Supplementary Information (see section 1). The method identified the best fit model based on the average of four inverted models, which was subsequently converted into water content. The optimized petrophysical parameters obtained are very close to the true values set for the study, suggesting that the petrophysical relationship is well constrained by the data. Validation dataset come from positions that are clearly separate from the training dataset locations. A good fit was found between the calculated and simulated water contents for both the training and validation datasets. Although the RMSE improvement from ensemble averaging was modest, the approach offers valuable insight into model uncertainty by quantifying the variability among individual inverted models. Based on several synthetic tests, we observed that the more heterogeneous the model, the more advantageous it becomes to combine multiple inverted models to minimize the RMSE. This likely stems from the inversion process struggling to adequately constrain complex or heterogeneous structures, making model combination a useful strategy to enhance robustness. These results demonstrate the method's applicability in a heterogeneous medium with two distinct petrophysical layers, a sharp vertical boundary in water content, and a vertical gradient.

## 3. Field applications

### 3.1. Field description

The EA-ERT methodology was applied on two real datasets from dissimilar sites. The two sites are located in southern France (Figure 2A): one in Avignon and the other on the Larzac plateau. Both sites have flat topography and a similar climate but differ markedly in terms of soil type and vegetation. The Avignon site is an experimental agricultural field that belongs to INRAE (French national research institute for agriculture, food and the environment) and is dedicated to studies of the soil plant atmosphere continuum in an agricultural setting. The soil is a relatively homogeneous calcareous silty clay loam to a depth of 0-60 cm, clay loam between 70 and 160 cm, and sandy loam below (> 160 cm). At the time of experiments, wheat was growing at the Avignon site. The Larzac site is a Mediterranean environmental research observatory (OREME, http://www.oreme.org/observation/gek/) dedicated to hydrogeophysical monitoring and experiments. It is part of the national hydrogeological observation service (SNO H+) of the OZCAR critical zone research infrastructure (https://www.ozcar-ri.org/). The Larzac site is a meadow that is adjacent to a stand of black pine trees on a heterogeneous dolomitic soil atop a karstic system. The primary characteristics of the sites are summarized in Table 1.

*Table 1: Characteristics of the two sites studied.*

| Site | Avignon | Larzac |
|---|---|---|
| Location | 43°54'59.0"N 4°52'45.5"E | 43°58'11.8"N 3°13'17.8"E |
| Altitude (m) | 30 | 705 |
| Climate | Mediterranean | Mediterranean |
| Annual precipitation | 663 mm (1990-2023) | 872 mm (2014-2023) |
| Annual temperature | 14.8°C (1990-2023) | 10.5°C (2014-2023) |
| Lithology | Silty clay loam Calcosol on coarse alluvium | Dolomite Karstic system |
| Soil thickness | Thick ≈ 100 -240 cm | Thin ≈ 70 cm |
| Soil profil | 0-60 cm = silty clay loam; 60-160 cm = clay loam; 90-110 = stony layer; >160 cm = sandy loam (and alluvium) | 0-35 cm = humus; 35-70 = dolomitic sand Dolomite rock outcrops |
| Environment type | Agricultural | Natural |
| Vegetation | Wheat | Grass, Boxwood (Buxus), black pines (Pinus nigra) |

### 3.2. Soil sampling and ERT measurement

Both sites are equipped with an ERT profile and soil sensors for measuring temperature and water content. ERT data acquisition was done using different measuring equipment and arrangements. The set-ups are shown in Figure 2. Data filtering and processing was not performed in the same way at both sites as reciprocal measurements were taken at Larzac but not at Avignon. Details of the ERT data acquisition and filtering are summarized in Table S2 in the Supplementary Information.

At the Avignon site (Figure 2B), 22 time-lapse measurements were repeated on the 25.2 m ERT profile between February and June 2016. The profile is divided into four equal areas characterized by different treatments of the wheat crop: S1) mowed - non-irrigated, S2) not mowed - non-irrigated, S3) not-mowed - irrigated, S4) mowed - irrigated. Each area was equipped with a neutron probe (Troxler) for soil water content measurement, thermocouples, and MPS sensors (Decagon) to measure soil temperature. Water content measurements were performed every 0.1 m from 0.15 to 1.45 m below the surface, and temperature measurements at 0.1, 0.3, 0.6 and 1.2 m. A Diver probe (VanEssen Instruments) was also used to obtain groundwater temperature measurements at a depth of 5 m. The neutron probes were calibrated using triple soil sampling and gravimetric measurements performed at each horizon under both dry and wet conditions. Four soil samples were collected at approximately ten meters from the ERT profile in the different soil horizons: 0-60, 60-90, 90-120 and 120-140 cm. Petrophysical relationships between electrical resistivity and saturation have been previously established in the laboratory on these samples as done in Doussan and Ruy (2009). Porosity and the parameters $F = \phi^{-m}$, n and $\sigma_s$ have been adjusted for each sample with Waxman and Smits (1968) equation. The values obtained are given in Table S5 in the Supplementary Information section and have been used as starting values for adjusting field relationships.

ERT surveys were carried out using an ABEM Terrameter SAS 4000, 64 electrodes spaced 0.4 m apart and left in place over time to acquire time-lapse ERT data. A dipole-dipole array including 922 quadrupole measurements was used. Quadrupoles with a stacking error greater than 3%, negative apparent resistivity values or values equal to 0 for a single time-step were removed from all time-lapse data sets to keep the same number of quadrupoles at each time step. After data filtering, the number of quadrupoles was 756, i.e. 166 quadrupoles were rejected.

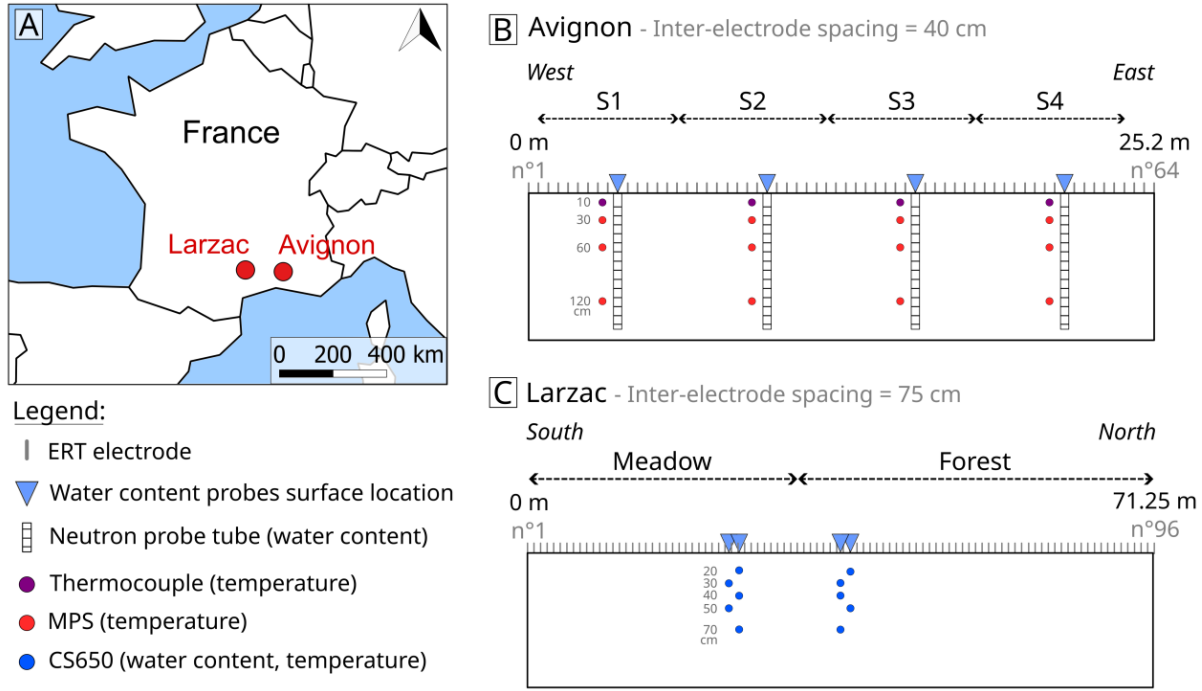

*Figure 2: A) Site locations in France and field configurations at B) Avignon and C) Larzac sites.*

At the Larzac site (Figure 2C), 13 time-lapse measurements were repeated over the 71.25 m ERT profile between April 2022 and September 2023. Two land uses are present at the site: a meadow and a black pine stand. Sensors to measure temperature and water content (Campbell CS650) were installed in each zone at 0.2, 0.3, 0.4, 0.5 and 0.7 m below the surface. The sensors were not specifically calibrated for the site; the Topp et al. (1980) equation was used to convert dielectric permittivity into volumetric water content. Water content data from the sensors was not corrected for temperature effect. Soil samples from each pit in which the sensors were installed and a dolomite sample were collected for laboratory measurements of electrical resistivity as a function of saturation, as done in Doussan and Ruy (2009), see the Supplementary Information for more details (Text S2). The values obtained are given in Table S5 in the Supplementary Information section and have been used as starting values for adjusting field relationships. In addition, seismic refraction measurements were acquired on the ERT profile to identify the distinct soil/rock boundary along the profile (Test S3 and Figure S13), see the Supplementary Information for more details (Pasquet et al., 2022).

ERT surveys were carried out using a Syscal Pro Switch 48 with a Switch Pro 48 (IRIS Instruments), with 96 electrodes spaced 0.75 m apart and left in place over time to acquire time-lapse ERT data. A Wenner-Schlumberger array was used to be sensitive to horizontal structures. This arrangement is more robust than dipole-dipole in the difficult field conditions associated with karst (e.g., Dahlin and Zhou, 2004; Samouëlian et al., 2005). Measurements were done each time in normal and reciprocal modes for quality assessment. Reciprocal measurement involves swapping current and potential electrode pairs. Theoretically, normal and reciprocal measurements should be identical. Normal and reciprocal measurements each include 2277 quadrupoles. The reciprocal error *err* was calculated for each quadrupole for each measurement using equation (9), where $R_r$ and $R_n$ are the resistances of the reciprocal and normal measurements, respectively. The error was increased by a fixed constant of 2% corresponding to the modeling error estimate (e.g., Mwakanyamale et al., 2012). This error was used for the inversion of apparent resistivities.

$$err = \left| \frac{R_r}{R_n} - 1 \right| + 0.02 \qquad (9)$$

In addition, data with negative apparent resistivity values or values equal to 0 were removed. After filtering, the number of quadrupoles was 2273, i.e. four quadrupoles were rejected.

### 3.3. Results and interpretation
#### 3.3.1. Generate inverted models

The proposed methodology was applied to data sets from the Avignon and Larzac sites. The $N$ inverted models were generated (see STEP 1 in Figure 1) using the open-source inversion code BERT (Boundless Electrical Resistivity Tomography; Günther and Rücker, 2015). BERT, like other inversion codes, is easy to automate and can manage many classic inversion parameters in addition to those associated with time-lapse inversions. The code uses a regularized Gauss-Newton approach based on finite element modeling techniques, a commonly used approach allowing broad flexibility in mesh construction (Günther et al., 2006). During the inversion process, BERT calculates the sensitivity matrix at each iteration, sums its column to derive the cumulative sensitivity (also called coverage), so as to obtain an indicator for assessing the reliability of the final inverted model.

The $N$ sets of inversion parameters used were selected based on the work of Audebert et al. (2014) and *a priori* knowledge of the environment. The values cover a realistic range while limiting the number of parameter sets. Both "*L1*" and "*L2*" norms were tested. Three "zweight" values were chosen to focus on horizontal structures only (0.01, 0.1 and 1). Values of "*λ*" were selected between 1 and 200 (1, 10, 50, 100, 200). In the case of BERT, the starting model for independent inversion is based on the median apparent resistivity of the dataset. By default, this starting model is also used as the reference model for the inversion. The ERT data sets were inverted independently and with two time-lapse strategies: i) with a common reference model, and ii) using the preceding time step model as reference model for the new time step. Two common reference models were tested: the model obtained from the first time step and the median of all independently inverted models (various time steps and inversion parameter combined), which is more representative of all time steps. When using the first time step model as a reference model, the value of "*λ*" is allowed to vary across time-steps, resulting in 25 combinations (5*5) related only to this parameter and therefore 150 inversions for this reference model (Table S1 in the Supplementary Information). The values tested for these parameters are summarized in Table S1, and the resulting number of inversions per type of reference model used is given. This resulted in a set of $N = 240$ inversion parameters, and consequently 240 time-lapse inverted models.

Of the 240 inversion parameter sets (Table S1), 10 inversions did not converge at Avignon, and nine did not at Larzac. Artifacts accumulated during the inversion of the various time steps, generating extreme and aberrant electrical resistivity values that prevented convergence. Parameters used for these inversions are listed in Table S3 in the Supplementary Information. Eight inversion parameter sets which have not converged are common to both sites. For all parameter sets that did not allow the inversion process to converge, the "L1" norm (sharp boundaries) is used and a time-lapse inversion is performed with the previous model as the reference. When the inversion of a time step did not converge with a parameter set, all time steps inverted with this parameter set were excluded from the analysis. As a result, 230 inverted models remain available for analysis at each time step of the Avignon time-lapse and 231 for the Larzac time-lapse.

### 3.3.2. Calibrate petrophysical relationship

To extract resistivity values at the water content probes locations for each of the 240 inverted models (see STEP 2 in Figure 1), the entire method was applied for several integration widths to identify the one minimizing the difference between inverted electrical resistivity converted to estimated water content and measured water content. At Avignon, the integration width for extracting electrical resistivity values at the water content probes cannot exceed 5 m to be able to integrate data from the same treatment (see the different treatments described in section 3.2 and visualized in Figure 2). An optimum width of 4.2 m was selected and used, i.e. 2.1 m on either side of the probe location. In contrast, the Larzac site is more heterogeneous, and an optimum integration width of 2.8 m was used. These integration widths were determined by running the method for widths ranging from 1 to 5 m in 0.2 m increments, and selecting the width that minimized the difference between the calculated water content (obtained from the inverted electrical resistivity) and the measured water content.

As the water content dataset is limited for the Larzac site, a strategy was adopted to split the data that does not yield fully independent subsets. For each time-lapse inverted model, the dataset (water content versus electrical resistivity) is sorted according to measured water content values and then divided into two equal subsets by selecting every other value. Each subset covers the same range of volumetric water content values. Inverted electrical resistivities collected on the meadow side during the August drought in the Larzac were excluded from further processing; since the data was low quality due to poor contact resistance (from ten to several hundreds of kΩ), it was decided not to use it for fitting the inverted electrical resistivity versus water content relationship. For each inverted model, 1210 and 87 inverted electrical resistivity values were extracted as functions of volumetric water content for the Avignon and Larzac sites, respectively. A clear relationship between the two variables is evident: electrical resistivity decreases with increasing water content. The data sets were equally divided, with 605 data in the two subsets for the Avignon site and 44 and 43 data respectively for the training and validation datasets for the Larzac site.

Equation 2 was optimized to determine the parameters $m$, $n$, $\sigma_w$ and $\sigma_s$ for the electrical resistivity/water content relationship on each time-lapse inverted model using the training dataset. The results from laboratory measurements on soil samples were used as starting parameters (displayed in blue in Figure S5 for Avignon and Figure S6 for Larzac, and noted in Table S5). Each parameter was constrained within the following bounds: $m \in [1.3, 5]$, $n \in [1, 5]$, $\log10(\sigma_w) \in [-2, -1]$ and $\log10(\sigma_s) \in [-5, 0]$ from Friedman, 2005 and Doussan and Ruy, 2009. The bounds provided here, particularly for $\log_{10}(\sigma_w)$ and $\log_{10}(\sigma_s)$, were chosen for our field sites. The petrophysical relationships were used to convert the electrical resistivity values into water content. The RMSE between the measured and calculated water content was then computed.

### 3.3.3. Inverted model averaging

The 240 inverted models were ranked based on the RMSE calculated for each model. Ranking the individual inverted models (see STEP 3 in Figure 1). Ranking the individual time-lapse inverted models makes it possible to identify the models yielding the lowest RMSE: parameter set n°156 for the Avignon site (3.59 % vol.) and parameter set n°126 for the Larzac site (2.44 % vol.) (see Table S4 in the Supplementary Information). The parameter set for Avignon uses the "*L2*" norm (smooth boundaries) and inverts the different time steps using the preceding model as reference. That for Larzac uses the "*L1*" norm (sharp boundaries) and inverts each time step using the first time step inverted model as the reference model. The regularization parameter "$\lambda$" is 1, a low value, for both sites. The anisotropy factor "*zweight*" is 0.1 for Avignon and is 1 for Larzac corresponding to isotropic regularization.

The inverted models were averaged to create ensemble models (see STEP 3 in Figure 1), starting with the highest-ranked inverted model (n°156 for Avignon and n°126 for Larzac) and successively adding each subsequent model according to the established ranking. A total of 230 ensemble models were obtained for Avignon and 231 for the Larzac, as some inversion parameter sets did not converge (see Section **Erreur ! Source du renvoi introuvable.**).

### 3.3.4. Recalibrate petrophysical relationship

New petrophysical relationships were fitted to each ensemble model in the same way as to the individual inverted models (see STEP 4 in Figure 1 and Section 3.3.2). After converting each electrical resistivity value from the ensemble model into water content, the RMSEs between the measured and calculated water content was calculated.

### 3.3.5. Model selection and assessment

Figure 3 shows the RMSE obtained for each ensemble model for both sites. The RMSE of the first ranked ensemble model ($r = 1$) is shown on the left of the graph, while the RMSE of all inverted models averaged together appears on the right, corresponding to the last ranked ensemble model ($r = N$, and $N = 230$ for Avignon and $N = 231$ for Larzac). The RMSE of the highest-ranked inverted model is represented by a green line in Figure 3 to visualize the evolution of the RMSE as individual inverted models are combined. For both sites, the result shows an improved RMSE for the first averaged models, reaching a minimum and then gradually increasing. The minimum RMSEs are reached when the eight highest-ranked inverted models are averaged at Avignon (Figure 3A) and when the six highest-ranked inverted models are averaged at Larzac (Figure 3B), as indicated by a green circle in Figure 3. These models were therefore selected as the ensemble models for each site (see STEP 5 in Figure 1). These ensemble models represent the best fit with field data of all the models evaluated.

Both "*L1*" and "*L2*" norm are used almost equally in the ensemble model selected for Avignon and "L1" norm is predominantly used for Larzac (see Table S4 in the Supplementary Information). The predominance of "*L1*" norm in the Larzac ensemble model likely reflects the site's greater heterogeneity, characteristic of karst systems, in contrast to the more homogeneous alluvial context of the Avignon site. The ensemble model selected for Larzac includes a majority of inverted models using the first time step inverted model, corresponding to relatively wet conditions, as a common reference model. Whereas the ensemble model for Avignon includes a majority of inverted models using the median model as a reference model. The Larzac model does not include independent inversions, while the Avignon model includes incorporates all the different inversion strategies. Lambda "$\lambda$" values are mainly higher in the selected inversions for the Avignon site, which may be explained by the site's homogeneous nature, where limited variability is expected and smoother models are appropriate. In contrast, at the Larzac site, characterized by strong heterogeneity and localized contrasts, lower lambda values are used to better resolve subsurface structures. The two models selected for the two sites do not include any common inversion parameter sets.

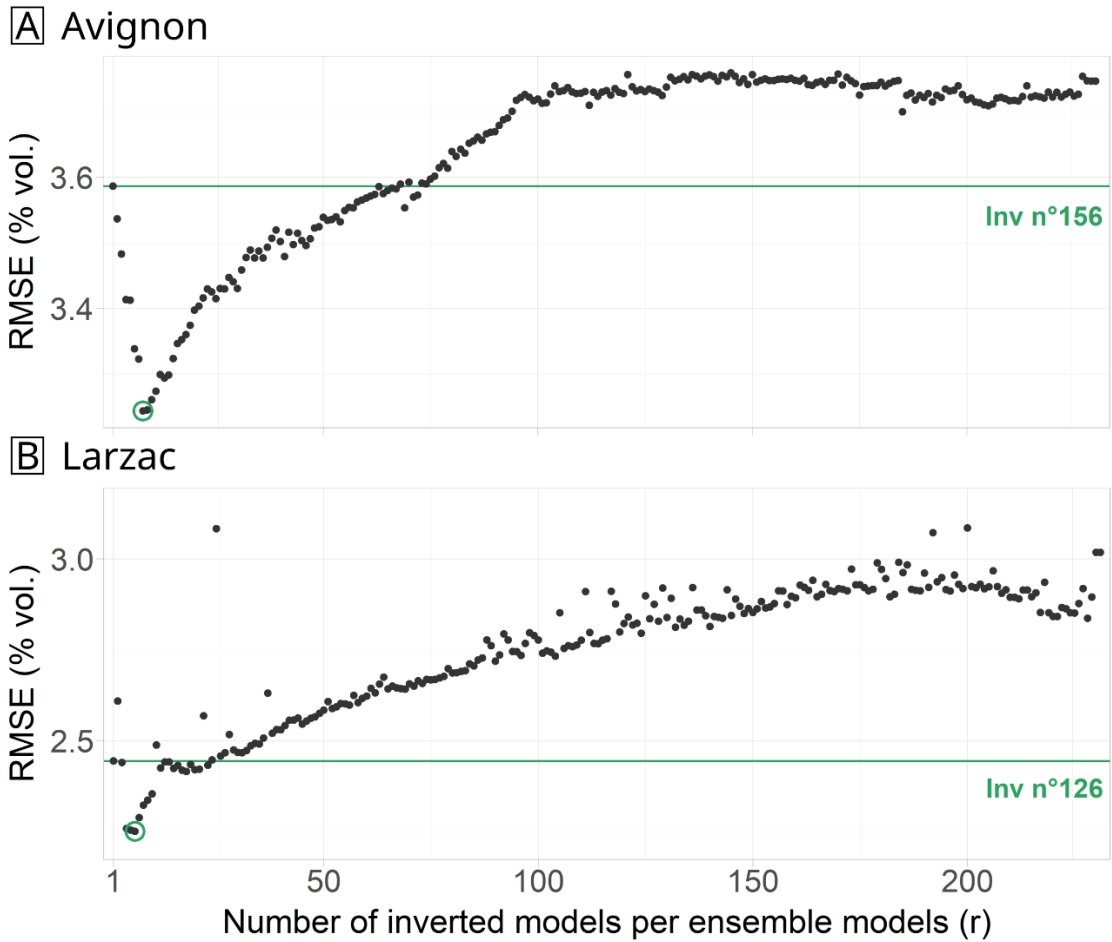

*Figure 3: RMSE results obtained between volumetric water contents measured with soil probes and calculated with ERT for each ensemble model A) at the Avignon site and B) at the Larzac site. The RMSE of the ensemble model including one inverted model, corresponding to the highest-ranked single model, is plotted on the left. The RMSE of the ensemble model including all inverted models (230 for Avignon, 231 for Larzac) is plotted on the right. The green line corresponds to the RMSE value of the single highest-ranked inverted model, i.e. the individual model with the lowest RMSE. This line provides a visual representation of the RMSE evolution as inversions are added to the averaged that forms the ensemble model. The green circle corresponds to the ensemble model with the lowest RMSE value and therefore to the selected model.*

Histograms of optimized parameter to fit the petrophysical relationships between electrical resistivity and water content for all ensemble models are displayed in Figure 4. This figure shows the values obtained for each parameter (m, n, $\sigma_w$, $\sigma_s$) for the first two soil horizons in Avignon (Figure 4A) and the two horizons under meadow in Larzac (Figure 4B). Histograms of optimized parameter values obtained for the other horizons are shown in the Supplementary Information (Figure S5 for Avignon and S6 for Larzac). In blue, overlaid on these histograms, are noted the laboratory-optimized parameters from soil samples which served as initial inputs for the field-scale optimization. In green are noted the optimized parameters obtained for the selected ensemble model. These histograms for each of the four optimized parameters show that the majority of the optimized values across the different ensemble models cluster around a central value. In Larzac, the selected optimized values lie close to the peak of the distribution, while in Avignon, they tend to be located further from it. This observation may suggest the added value of an ensemble approach, as the optimal solution does not necessarily align with the most frequently occurring parameter values.

The values obtained for these parameters are consistent with the known properties of the sites. At Avignon, characterized by silty clay loam soil, the porosity values used for this site, ranging from 0.38 to 0.43, are consistent with the value reported by Doussan and Ruy (2009) at the same site, which was 0.38. Our results align with those of Doussan & Ruy (2009) and Ghorbani et al. (2008), showing high surface conductivity (between 0.158 and 0.274 S/m) and water conductivity slightly below 0.1 S/m. At Larzac, where dolomitic sand overlays dolomite, porosity obtained in the laboratory and used in the adjustment reflect literature values: 0.45–0.49 for sand (e.g., Friedman 2005) and 0.095 for dolomite consistent with Fores (2016) at the same site. Water conductivities (0.01 to 0.03 S/m) match continuous borehole measurements (up to 0.04 S/m). Lower surface conductivities (≤ 0.026 S/m) are not directly referenced in the literature for this soil type but are comparable to those of Fontainebleau sand in Doussan & Ruy (2009) with a value of 0.036 S/m. Finally, the values of the cementation (m ≈ 2) and saturation (n ≈ 2-3) exponents obtained at both sites are consistent with commonly accepted ranges (e.g., Glover, 2015).

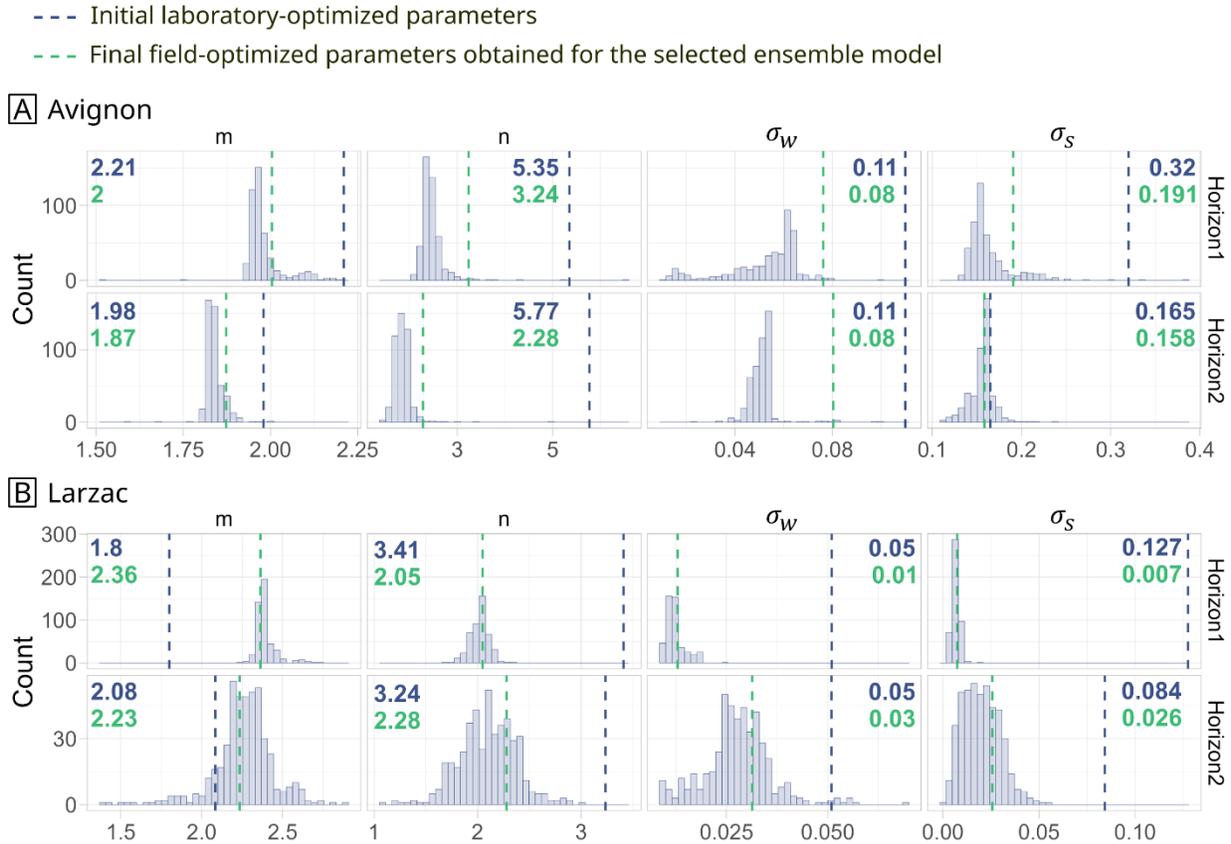

*Figure 4: Histograms of the four optimized petrophysical parameters (m, n, $\sigma_w$, $\sigma_s$) with the Waxman and Smits (1968) electrical resistivity-water content relationship for the first two soil horizons for each of the ensemble models A) at the Avignon site and B) at the Larzac site (equivalent to the two horizons under the meadow). The blue dashed lines and their corresponding values show laboratory-optimized parameters used as the initial values for the field-scale optimization. The green lines and their corresponding values show field-optimized parameters obtained for the selected ensemble model.*

Figure 5 presents the comparison between water content calculated from the selected ensemble models and measured from field sensors with the training and validation dataset for both sites. The high correlation coefficients ($r$ = 0.8 and 0.94, respectively for Avignon and for Larzac) with very low p-values (< 0.001) indicate significant and strong relationships between the calculated and observed data. The low RMSE values of 3.24% vol. and 2.25% vol., respectively for each site, suggest a good quality of fit. Evaluation of the selected models on the validation dataset also shows strong correlation coefficient ($r$ = 0.81 and 0.95, respectively) between measured and calculated water content data, similar to those obtained on the training dataset. The RMSEs obtained are 3.11% vol. and 2.23% vol. for the Avignon and Larzac sites, respectively. These results apply to both sites, suggesting no obvious overfitting. However, it may also reflect the sampling strategy, which is not ideal since the training and validation datasets are not completely independent. In Figure 5, the error bars represent the weighted standard deviation of the averaged electrical resistivities from the different models composing the ensemble. For the Avignon site (Figure 5A), the error bars that appear saturated actually correspond to relatively low average electrical resistivities (around 100 Ω.m), but with a large standard deviation, on the order of the mean value itself. This results in minimum resistivity values close to 0 Ω.m, which cannot be converted into water content because they fall outside the range defined by the calibrated petrophysical relationships. In such cases, the maximum water content is manually set to the porosity value of the corresponding soil horizon.

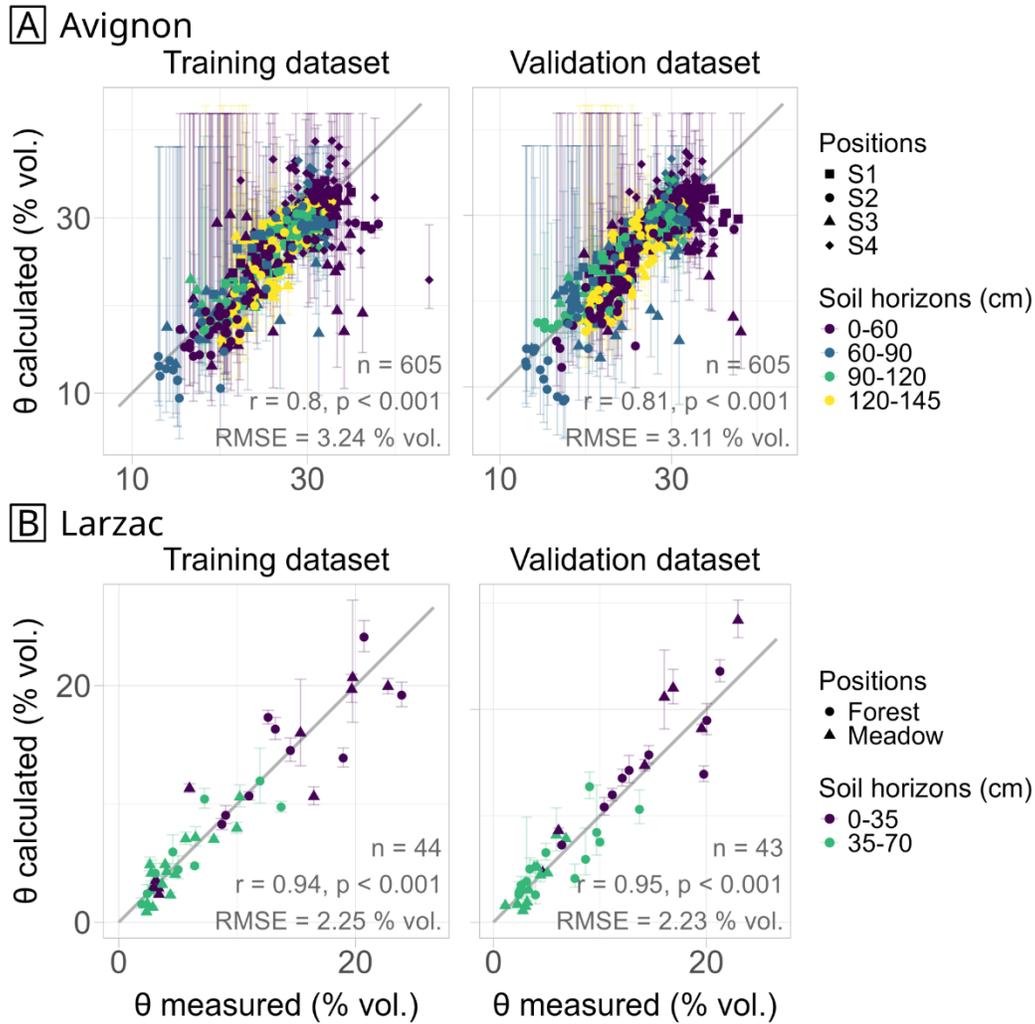

*Figure 5: Volumetric water content θ (% vol.) measured with soil water content probes compared to those calculated with ERT for the selected inversion model A) at the Avignon site and B) at the Larzac site. On the left is the correlation obtained on the training dataset and on the right is the evaluation made on the validation dataset (RMSE: root mean square error; r: correlation coefficient; p: p-value of r, n: number of value).*

Figure 6 shows ERT sections converted to volumetric water content using the ensemble model selected for each site (see Figure S7 in sup. mat. to see the same ERT sections in electrical resistivity). Resistivities were converted using equation 2 with the parameters obtained for each soil horizon (noted in green in Figure 4). The petrophysical relationships adjusted for the deepest soil horizon were used to convert electrical resistivities below this horizon for the Avignon site. At the Larzac site, the boundary between soil and rock was identified using seismic refraction (Figure S13). The laboratory-adjusted petrophysical relationship for the dolomite sample was used to convert electrical resistivity to water content in the rock section (see parameter values in Table S5) since no direct measurements of water content were available. Discontinuities are noticeable in Figure 6B due to the different petrophysical relationships used (soil/rock and meadow/forest). Only the first few meters below the surface are shown, partly because the models were calibrated on water content data down to 1.45 m for the Avignon site and 0.7 m for Larzac. Additionally, we extend the display slightly deeper to highlight areas identified as poorly constrained by the inversion process, particularly where the sensitivity coverage drops below zero.

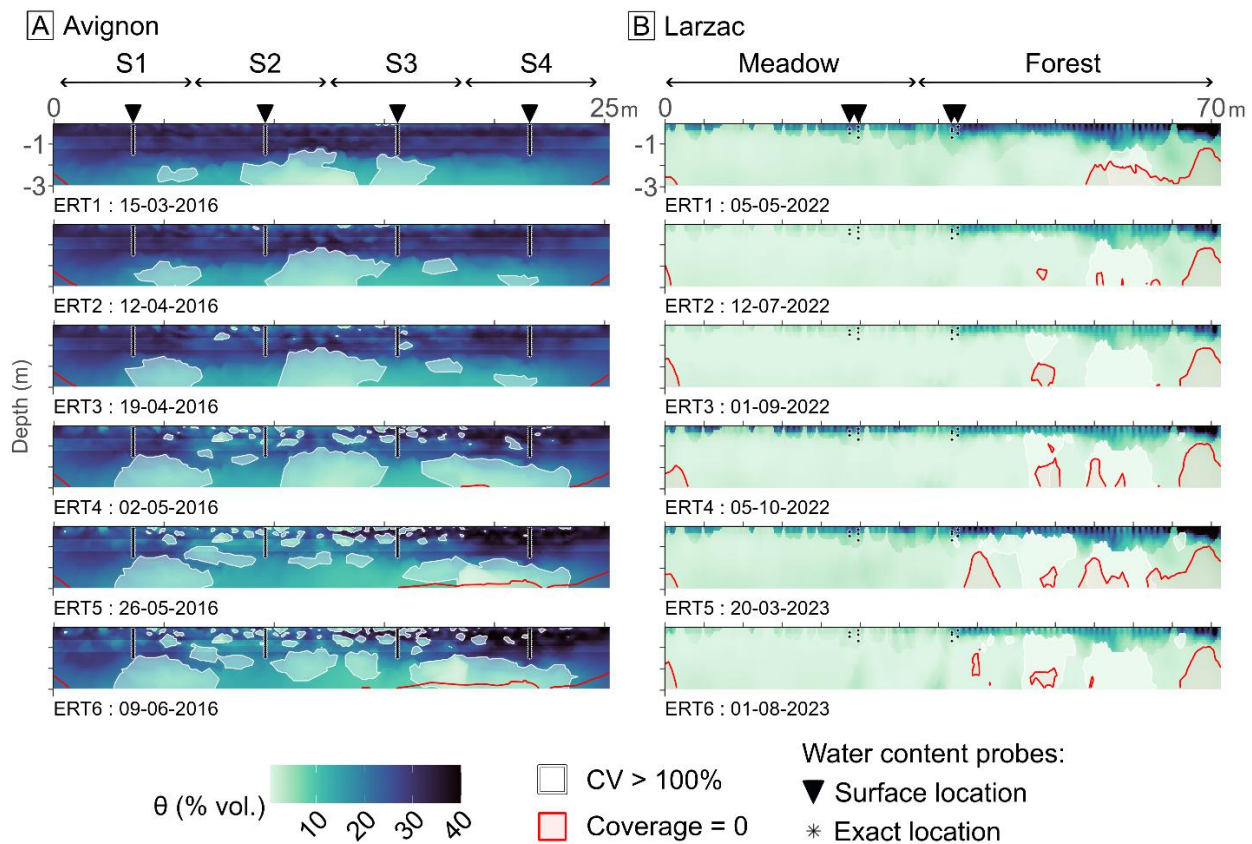

*Figure 6: Example of six time steps of 2D ERT section converted to volumetric water content θ (% vol.) (with best fit parameters noted in green in Figure 4) A) at Avignon and B) at Larzac. Slightly transparent white areas correspond to areas where the coefficient of variation (CV) exceeds 100%.* The red ones corresponds to areas where the coverage is equal to or less than 0.

Two horizons seem to stand out on the wetter early time steps of the Avignon site, the first from the surface to a depth of 1-2 m, with a high water content, and the second at depth, with a slightly lower water content (Figure 6A). The soil dries out as spring progresses, and differences in water content changes between the different treatments can be observed. Water content decreases most noticeably in the S2 and S3 areas in the middle of the section, corresponding to the unmowed areas, where vegetation is more abundant and therefore extracts more water. The S3 and S4 areas have a wetter soil surface than the others, consistent with irrigation in these two areas. The S4 area has a higher water content than the others, as it is irrigated and mowed. The Larzac site is characterized by a very low overall water content (Figure 6B), especially in the dolomite part. Soil is much thicker under the trees (forest part on right) than under the meadow (part on left), where dolomite outcrops in places. The soil water content under the trees is higher than under the meadow at all times, which we attribute to the greater soil thickness. Soil water content remains relatively high even in the middle of summer (ERT2, ERT3 and ERT6), in contrast to the meadow side, which dries out.

The red line represents the contour line where coverage is equal to 0 (in logarithm base 10). Coverage is the indicator provided by the BERT inversion code for assessing the reliability of the final inverted model (see Section **Erreur ! Source du renvoi introuvable.**). Here, it corresponds to the coverage-weighted average obtained for each model that compose the selected ensemble model. Below this value, areas are considered to be insufficiently covered by the data and therefore poorly constrained by the inversion process. In the literature, it is often challenging to define a precise threshold for such indicators (e.g., Robert et al., 2012; Caterina et al. 2013). In our case, the value was set subjectively, but complementary analyses or measurements would be valuable to better constrain this threshold. Coverage is computed for each inverted model during inversion process. For the ensemble model, coverage was obtained by weighted averaging of the coverage values from the individual models that were combined. The coverage line is fairly regular at the Avignon site, unlike at Larzac. The line is located between 3 and 4 m below the surface over the entire profile at Avignon, whereas at Larzac it descends to depths of over 5 m below the meadow and rises to very near the surface below the trees.

The coefficient of variation (CV) between the electrical resistivities of the different averaged inverted models was calculated (see STEP 5 in Figure 1). The slightly transparent white areas represent zones where the *CV* exceeds 100%. These areas are considered to be highly uncertain. On both sites, this threshold is reached in areas where water content is low: at the boundary between the two horizons that have dissimilar water contents at Avignon, and in very desiccated halos under the trees on Larzac. The exact values of the *CV* across all sections are presented in Figure S8 in the Supplementary Information. A high *CV* occurs when the electrical resistivity is extremely high for both sites beneath a highly conductive shallow layer, a known artifact resulting from the inversion process (e.g., Clément et al. 2009). This indicator highlights areas where the combined resistivity models diverge (possible artifacts), potentially due to insufficient data coverage or low sensitivity in certain parts of the model, making the interpretation of data in these zones challenging. These white zones appear closer to the surface than the coverage threshold line in Avignon and suggest that only the values from the first horizon (up to 1-2 m depth) with a higher water content can be interpreted reliably. At Larzac, the white zones coincide with areas where the coverage threshold line is present and rises closer to the surface. The combine use of CV and coverage appears to be complementary. For example, at Larzac, both indicators highlights numerous unreliable zones beneath the trees, below the wetter surface horizon, which supports the idea that resistivity and water content values should not be interpreted in these areas.

## 4. Discussion

In this paper we propose the Ensemble Approach ERT (EA-ERT) method to facilitate ERT data processing with the aim of obtaining spatiotemporal water content variations. This method relies on ERT data, point soil water content measurements and petrophysical functions. The method responds to the issues presented in the introduction: i) choice of inversion parameters; ii) assessment of ERT model reliability; and iii) conversion of the geophysical signal to water content. The applications using one synthetic data set and two field data sets demonstrate the applicability of this methodology under dissimilar various conditions. In the following paragraphs, we discuss the three points mentioned above.

First, the choice of inversion parameters is sometimes difficult, as it requires in-depth knowledge of geophysical data processing. Incorrect parameter choices can significantly affect final interpretations. The EA-ERT method circumvents this problem by generating and evaluating a large number of inverted models (240 in our case) obtained by using different parameter sets (see step 1 in Figure 1). Four parameters were tested, based on the work of Audebert et al. (2014): the model norm, the anisotropy factor, the regularization parameter and the reference model. It is also possible to include additional parameters to obtain a greater number of models (e.g., the starting model) or a larger number of values for each parameter. We selected only five values for the regularization parameter « λ » (Table S1), to cover a fairly wide range while limiting the number of parameters sets to minimize computation time. Obviously, depending on the study site and prior knowledge, some parameters may be more or less relevant. For example, the inversion that used the previous reference model in time-lapse seemed to be less relevant. In fact, one third of these inversions were unsuccessful due to the accumulation of artifacts (i.e. extreme values) during the most advanced time steps, and very few of the inversions were included in the final ensemble models (Table S3). The procedure for evaluating the generated inversions (step 2 and 3 in Figure 1) facilitates the user's interpretation by identifying the most effective models for estimating water content. This selection of best models is then used to calculate an ensemble model whose performance exceeds that of the best individual inverted model.

The various available inversion software packages do not use exactly the same inversion parameters. Nevertheless, the method could be applied in a similar way, by identifying the parameters that most affect the inverted model in the used software, and varying them to generate a large number of models.

Second, assessing ERT model reliability is essential for interpretation and decision-making. Inversion software automatically calculates one or more reliability indicators (e.g., coverage, sensitivity, resolution, DOI), which are useful for assessing depth of investigation (e.g., Caterina et al., 2013, Paepen et al., 2022) but they are not designed to detect localized artifacts and do not fully capture spatial uncertainty. The ensemble approach proposed in the EA-ERT method not only provides a more efficient model for predicting water content, it also reveals a form of uncertainty that is represented by the dispersion between the models used to make the ensemble model. This dispersion is linked to the non-uniqueness of the inversion process, depending on the parameters chosen. Reliability of the final ensemble model was assessed using the coefficient of variation proposed by Vinciguerra et al. (2024). This indicator highlights areas where dispersion between inversions is high, similar to the DOI (Oldenburg and Li, 1999), but with an ensemble of models instead of two. For both sites, the CV highlights the difficulty of describing resistant zones (low water content) under a conductive body (high water content) as can be seen on the electrical resistivity panels in the Supplementary Information (see Figure S7 and S8) and on the water content panels (Figure 6). This result confirms the artifacts that Clément et al. (2009) highlighted in similar configurations by direct modelling. In their study, they demonstrate that when electrical resistivity decreases near the surface, inversion algorithms tend to overestimate resistivity values at depth.

Third, converting the geophysical signal into a directly useful parameter such as water content is quite complex, because the geophysical signal is affected by multiple parameters (e.g., temperature, electrical conductivity of water, clay content, salinity; see Friedman, 2005; Samouëlian et al., 2005; Glover 2015). This is particularly true in field studies. In the literature, petrophysical relationships are often calibrated on soil samples in the laboratory, where electrical resistivity is measured at different saturations (e.g., Doussan and Ruy 2009; Laloy et al., 2011; Brillante et al., 2015; Dimech et al., 2023). In a preliminary phase of this work, inverted models from the Larzac site were converted to water content using this laboratory-based approach. While a correlation was observed (r = 0.85, p-value < 0.001) with a RMSE a relatively small (3.72% vol.), the results were unsatisfactory compared to the approach based on direct calibration using field data: water contents were largely underestimated at higher values and tended to be overestimated at lower values (see Figure S9). The sampled soil is generally disturbed during the extraction process, making it unrepresentative of in-situ conditions. In addition, such laboratory measurements often involve problematic scale mismatches compared to field-scale observations. As a result, translating homogeneous laboratory data into more heterogeneous field conditions remains a well-known challenge, as noted in several studies (e.g., Michot et al. 2003; Brillante et al., 2015; Tso et al. 2019; Dimech et al., 2023). The EA-ERT method addresses this issue by directly calibrating inverted models using field data. This approach is more robust, but it also requires more time to collect a sufficient variation in water content to accurately constrain petrophysical relationships (e.g., Brillante et al. 2015). However, this can be challenging depending on the study sites; for example, at Larzac, water content is often very low, limiting the range of values available for calibration.

Field-scale calibration of petrophysical relationships raises important challenges, particularly concerning the direct comparison between point water content measurements and the integrative measurement of ERT, a topic already discussed in several studies (e.g., Koestel et al. 2008; Brillante et al. 2014; Benoit et al. 2019; Dimech et al. 2023). Successful inversion calibration using field data depends on sensor's depth and location as well as on the setup used to perform ERT measurements. Specifically, the depth of investigation and resolution of ERT measurements are controlled by the electrode spacing and the support volume of the inversion mesh. To capture resistivity in shallow soil horizons, a narrow electrode spacing (<1m) is essential. If the spacing is too wide, the ERT signal may not be sensitive to the zone where water content sensors are located near the surface, which would compromise calibration accuracy.

The choice of ERT integration volume is crucial: it must be large enough for the extracted resistivity value to be representative and homogeneous with respect to the surrounding water content, while remaining consistent with the spatial resolution of the ERT model. Integration volume must be determined so that resistivity values can reliably reflect the measured water content. In our work, we deliberately selected an integration window centred on each sensor location. Several horizontal averaging widths were tested to smooth out lateral heterogeneities, but this had a limited effect on the results. However, this choice is not ideal, as the integration volume does not account for the vertical variability in resistivity. It would also be relevant to adjust the integration height around each sensor.

It is also important to recognize ERT resolution decreases with depth, meaning that the effective volume of investigation increases. It may be advisable to integrate resistivity values over a volume that varies with depth. However, it introduces additional complexity. The integration volume is influenced not only by electrode spacing and array geometry, but also by the inversion mesh, inversion settings and the regularization applied (e.g. Benoit et al., 2019), and may therefore vary from one inversion model to another.

Ultimately, It is important to acknowledge that petrophysical relationships are inherently uncertain and cannot be considered entirely accurate or error-free (e.g., Binley et al., 2015; Tso et al., 2019). In our case, the conversion relies on relationships that include four unknown parameters (m, n, $\sigma_w$, $\sigma_s$) which are optimized, a non-negligible source of uncertainty. Additionally, these relationships are sensitive to the complex dependence of electrical resistivity on both interconnected pore fluids and interconnected surface conduction pathways (e.g., Weller et al., 2013), further contributing to the uncertainty.

The EA-ERT method is a novel approach compared to conventional ERT data processing. This study is exploratory, and we are acknowledge that the method relies on certain assumptions that may be subject to discussion. Integrating direct field measurements into ERT data analysis provides a resistivity model that is consistent with field data. However, this approach raises fundamental questions about the homogeneity of the environment in terms of the measured property (water content) and its sampling. The number of measuring points at Avignon was unusually large (four profiles down to 1.5 m over a section of just 25 m). Nevertheless, by fitting the petrophysical relationships to the few field measurements available and applying them to the entire section, we assume the environment to be homogeneous in terms of properties such as texture or soil type. This assumption is a known limitation since natural soils are inherently heterogeneous at various scales.

In this approach, water content measurements obtained from various sensors (e.g., TDR, FDR, neutron probes) are treated as ground truth. However, these measurements also carry inherent uncertainties, which were not accounted for in this study. These uncertainties stem from sensor calibration, installation procedures, and measurement errors (e.g., S.U. et al., 2014). Nonetheless, despite these limitations, we consider the sensor data reliable enough to serve as a reference for evaluating and converting ERT-derived resistivity data into water content estimates. For example, according to the manufacturer, the TDR sensors (CS650, Campbell) installed at the Larzac site have an accuracy of approximately ± 3% and a precision better than 0.05%. Our main objective is not to assess the absolute accuracy of the measurements, but rather to propose and demonstrate a method for processing and converting ERT data into soil water content estimates.

The EA-ERT method we propose ideally relies on relatively large datasets to be applied as rigorously and reliably as possible. However, the limited size of the available datasets, particularly for the Larzac site, forced us to adopt a data-splitting strategy for training and validation that is not optimal as the two subsets are not fully independent. In particular, it did not allow for a more distinct partitioning, such as reserving certain sensors or time periods exclusively for validation. To overcome this limitation and ensure a meaningful validation of the method, we conducted a synthetic study. This analysis highlights the added value of the ensemble approach, especially in contexts where the subsurface model is heterogeneous. In such cases, the inversion process tends to struggle to constrain the model adequately. Combining multiple inverted models thus helps improve the robustness of the final estimate and provides valuable insight into spatial uncertainty.

## 5. Conclusions

In this paper, we propose a new method called the Ensemble Approach ERT (EA-ERT) to derive water content and associated uncertainties from ERT measurements. This ensemble approach allows us to: i) circumvent the ERT inversion parameterization problem by evaluating the performance of a large number of models calculated with different parameter settings; ii) provide an uncertainty estimate of the final model represented by the dispersion between models; iii) converts ERT data to water content from field measurements. The approach integrates point-scale water content measurements to calibrate ERT data quantitatively. The objective is to obtain a 2D subsurface water content distribution along the ERT profile in areas lacking probes, enabling the calculation of distributed water volumes. Our method uses time-lapse ERT measurements to provide sufficient data under different humidity conditions to properly fit petrophysical relationships.

The method analyzes and combines multiple inversions outputs through a five-step process: 1) a series of resistivity models is generated from time-lapse apparent resistivity data using a range of inversion parameter sets; 2) each model is then evaluated by converting resistivity values into water content using petrophysical relationships and comparing the results to in-situ probe measurements. The root mean square error (RMSE) between calculated and measured water content is used as the evaluation criterion; 3) based on this evaluation, a set of ensemble models is constructed by computing weighted averages of the resistivity models, where the weights reflect each model's performance; 4) these ensemble models are then evaluated using the same procedure as for the individual inverted models; 5) the ensemble that best aligns with the field data is selected, and its reliability is assessed by calculating the coefficient of variation among the averaged models it contains.

The EA-ERT method was applied to a synthetic data set and two data sets from dissimilar field sites. For each case study, building ensemble models based on the RMSE of the individual inverted models resulted in a lower overall RMSE. The electrical resistivity values of the ensemble models were converted into water content, achieving a good fit with probe-based measurements (3.24% vol. for Avignon and 2.25% vol. for Larzac). The selected ensemble model consistently included fewer than 10 averaged individual models. The same sets of parameters were used at each site to generate the inversions, but no individual inversion was shared between the selected ensemble models. Selected ensemble models primarily relied on time-lapse inversions using a common reference model for each dataset. Without introducing any prior assumptions, smoother models were ultimately selected for the Avignon site, while sharper, more contrasted models were favored for the Larzac site. These observations align well with our knowledge of the respective field contexts: a relatively homogeneous agricultural area versus a heterogeneous karst media. Combining multiple models allowed us to calculate the dispersion among them using the coefficient of variation, providing an uncertainty assessment. This represents the key advantage of the ensemble approach, beyond merely reducing the RMSE value which is mainly used to make a choice to select the final ensemble model. The results demonstrated the effectiveness of the EA-ERT method and validated its robustness.

The EA-ERT method proposed in this article is a novel approach compared to conventional ERT data processing and analysis and it opens up new perspectives. It is evident that this study is exploratory and preliminary, requiring further work on various aspects to enhance the method's robustness, particularly by reducing sources of uncertainty or improving their quantification. For example, issues related to the scale mismatch between the two measured and compared variables, the use of water content values as the "true" reference, and uncertainties in the petrophysical relationships need to be addressed. The method could be further improved by incorporating more advanced mathematical frameworks, such as Bayesian approaches, to better characterize and manage these uncertainties. In the long term, it would be highly valuable to develop a workflow that enables a user-friendly conversion of ERT data into water content, ideally making it accessible even to non-specialists. The method has the advantage of being easily reproducible and it can be applied to convert ERT data to other parameters of interest (e.g., temperature, electrical conductivity of water) in hydrology, plant science, ecology and Critical Zone science in general.

## Acknowledgments

The authors would like to express their gratitude to H+ hydrogeological site network (SNO H+, https://hplus.ore.fr/en/) of the OZCAR critical zone research infrastructure (https://www.ozcar-ri.org/), which is supported by the French Ministry of Research, French Research Institutions and Universities, and the OREME observatory network (http://www.oreme.org/) for providing access to the Larzac site. The authors strongly thank the financial support of ANR TAW-tree (grant ANR-23-CE01-0008) for this work. This study is founded by the French Ministry of Education and Research for a PhD grant.

We would also like to thank Marc Dumont and Frédéric Nguyen for the very interesting upstream discussions on this work, and Lise Durand for help in the field work.

## Open Research

The present study is supplemented by Supplementary Information. Field data on ERT, soil water content and soil temperature, as well as Waxman and Smits model parameters optimized in the laboratory on soil samples, are freely available to the hydrogeophysics community via Zenodo:

https://doi.org/10.5281/zenodo.13142111. In addition, the R codes for each step of the proposed method are also available via this link.